\documentclass[10pt, conference, compsocconf]{IEEEtran}

\IEEEoverridecommandlockouts

\usepackage{cite}

\usepackage{graphicx}
\DeclareGraphicsExtensions{.eps}
\usepackage{subcaption}

\usepackage{amsmath}
\usepackage{amsfonts}
\usepackage{bbm}

\usepackage{array}

\usepackage{fixltx2e}

\usepackage{url}

\begin{document}

\title{Consensus as a Nash Equilibrium of a
Dynamic Game}

\author{\IEEEauthorblockN{Muhammad Umar B. Niazi}
\IEEEauthorblockA{Department of Electrical and\\Electronics Engineering,\\
Bilkent University,\\
Ankara, Turkey.\\
Email: niazi@ee.bilkent.edu.tr}
\thanks{This work is supported by the Science and Research Council of Turkey (T\"{U}B\.{I}TAK) under the project EEEAG-114E270.}
\and
\IEEEauthorblockN{Arif B\"{u}lent \"{O}zg\"{u}ler}
\IEEEauthorblockA{Department of Electrical and\\Electronics Engineering,\\
Bilkent University,\\
Ankara, Turkey.\\
Email: ozguler@ee.bilkent.edu.tr}
\and
\IEEEauthorblockN{Aykut Y{\i}ld{\i}z}
\IEEEauthorblockA{Department of Electrical and\\Electronics Engineering,\\
Bilkent University,\\
Ankara, Turkey.\\
Email: ayildiz@ee.bilkent.edu.tr}
}

\maketitle

\begin{abstract}
Consensus formation in a social network is modeled by a dynamic game of a prescribed duration played by members of the network. Each member independently minimizes a cost function that represents his/her motive. An integral cost function penalizes a member's differences of opinion from the others as well as from his/her own initial opinion, weighted by influence and stubbornness parameters.  Each member uses its rate of change of opinion as a control input.  This defines a dynamic non-cooperative game that turns out to have a unique Nash equilibrium.  Analytic explicit expressions are derived  for the opinion trajectory of each member for two representative cases obtained by suitable assumptions on the graph topology of the network. These trajectories are then examined under different assumptions on the relative sizes of the influence and stubbornness parameters that appear in the cost functions.
\end{abstract}

% \fi
\IEEEpeerreviewmaketitle

\vspace{0.25cm}

\section{Introduction}

How gossip  spreads in a small community, how a political leader reaches or fails to reach voters, and how some students learn faster than others among those with comparable intellectual capacity are three questions that fall into the study of social opinion dynamics. It is no surprise that the research question has attracted the attention of many disciplines in a short span of time and a sizable penetrating literature has been accumulated. We refer to the survey papers \cite{ozdaglar}, \cite{albi} and \cite{olfati} for only a partial panorama. These publications can roughly be divided into those that take a Bayesian perspective such as \cite{welch} and those that put forward non-Bayesian models such as \cite{degroot}. Yet another classification is that while most of the research focuses on formation of a consensus \cite{tsitsiklis}, there are also those that study disagreement as in the case of Hegselmann and Krause model \cite{krause}, \cite{etesami} or as in \cite{bindel}. The study of consensus has several engineering applications including multi-agent coordination \cite{ren}, information fusion in sensor networks \cite{boyd}, consensus in small-world networks \cite{olfati2} and distributed optimization algorithms \cite{tsianos}.
 
We study consensus formation via Nash equilibrium in a dynamic game of a prescribed duration played by members in a social network. Each member (player or agent) independently minimizes a cost function that represents ``its" (can be read as ``his/her") motive. An integral cost function penalizes its differences of opinion from its neighbors as well as from its own initial opinion, weighted by influence and stubbornness parameters.  Each member uses its rate of change of opinion as a control input. 
This defines a dynamic non-cooperative game that turns out to have a unique Nash equilibrium. For two representative cases obtained by suitable assumptions on the information structure (graph topology), we are  able to obtain 
explicit analytic expressions for the opinion trajectories of all members in the Nash solution. These trajectories are then examined under different assumptions on the relative sizes of influence and stubbornness parameters.    
    
Nash equilibrium is only one among a wide range of equilibrium concepts in games.
One interpretation in  \cite{osborne} suggests that if the same game is played several times with no strategic dependencies between consecutive plays, then a Nash equilibrium is most likely reached. This is for static games but one can extend the interpretation to dynamic games as well. The point of the matter is that it is a very useful construct (and presently the only rigorous one) if the research objective is to examine under what conditions, from independent motives of agents, a pattern of collective behavior emerges.

In \cite{ghaderi}, a static game of opinion dynamics is posed and the best response function in a Nash solution is used to postulate an update scheme. The convergence of this dynamic scheme to a consensus is examined. One can view our game model here as a dynamic version of \cite{ghaderi}. The optimal control of consensus model and control through a leader model in \cite{albi} also use integral cost functions and has similarities to our model except that  the objective in their case is control of consensus via external actions. The non-cooperative dynamic game model here is inspired by the foraging biological swarm models in \cite{ozguler}, \cite{ozguler2}, and \cite{ozguler3}.

In the next section we pose the opinion dynamics game in its most generality. In Section 3, we study two specialized versions and obtain explicit Nash solutions for these two games that represent extreme cases of information structure. Section 4 contains a number of simulation results for the games of Section 2 and 3. The last section is on conclusions.

\section{A Game of Opinion Dynamics}

We represent a social network of $n$ agents by a weighted directed graph $G=( N , E, w_{ij})$, where $N =\lbrace1,...,n\rbrace$ is the set of all nodes (agents), $E  \subseteq N  \times N $ is the set of all ordered pairs of connected nodes, and $w_{ij}$ is the influence of agent $j$ on agent $i$ when $(i,j)\in E$. One-sided or two-sided connection between the nodes indicate one-sided or two-sided interaction between the agents. The neighborhood of agent $i$ is defined to be the set of all agents with whom agent $i$ interacts, i.e., $\eta_i := \{j\in N  : (i,j)\in E\}$. The reason for a directed graph representation is because we can interpret the weight on the edges to be the influence of an agent on its neighbor or the value its neighbor gives to the opinion of an agent. Thus, two neighbors can have different levels of influence on each other. Let $x^i(t)$ be the opinion at time $t$ of agent $i$ and let it be normalized so that for every $t$ in the interval $[0,T]$, $x^i(t)\in [0,1]$. Each agent has an initial opinion $x^i(0)=x_0^i\in [0,1]$ about a certain issue, where the values $0$ and $1$ indicate the extreme cases. For example, $0$ may be interpreted as strong disagreement and $1$ as strong agreement cases. Let $\mathbf{x}(t)=[x^1(t)\,...\,x^n(t)]'\in [0,1]^n$ denote the opinion profile at time $t$  in the network of $n$ agents, where `prime' denotes transpose. The cost functional of agent $i$ is postulated to be    
\begin{multline} \label{cost}
	L^i(\mathbf{x},x_0^i,u^i) = \int_0^T \biggl( \frac{1}{2} \sum_{j\in \eta_i} w_{ij} \left[ x^i(t)-x^j(t) \right]^2 \\
	+ \frac{1}{2} k_i \left[x^i(t)-x^i(0) \right]^2 + \frac{1}{2} \left[u^i(t)\right]^2 \biggr) dt,
\end{multline}
where $w_{ij}\in [0,\infty)$ is the parameter that weighs the susceptibility of agent $j$ to influence agent $i$, $k_i\in [0,\infty)$ weighs the stubbornness of agent $i$ or the reluctance of $i$ to divert from its initial opinion. The control of agent $i$ is assumed to be $u^i(t)=\dot{x}^i(t)$, so that agent $i$ controls the rate of change of its opinion. The coefficient of the control term in the cost is normalized to $1$, without loss of generality. The integral in the time interval $[0,T]$ indicates that the agent penalizes the cumulative effect in each of the three terms in the integrand. Considering the first term, for instance, what it penalizes as part of the cost is the sum total of the divergence from the opinions of the neighbors, not the instantaneous differences from their opinions. This cost functional, which should be viewed to be a model of the motive  of agent $i$ towards a prevailing social issue, is prompted by \cite{ghaderi}, in which a static model for the motives of agents in a social network was used and by \cite{ozguler}, in which a similar cost functional modeled the motives of members in a foraging biological swarm. If each agent in the social network minimizes its cost, then we have a non-cooperative dynamic (or, differential) game played by $n$ agents 
\begin{equation} \label{game}
	\min_{u^i} \lbrace L^i \rbrace \mbox{ subject to } \dot{x}^i(t) = u^i(t) \mbox{ \ \ \ } \forall i\in N. 
\end{equation}
A solution to such a game, if it exists, is a {\it Nash solution}, or a {\it Nash equilibrium} of the game.
Note that although $\mathbf{x}(0)$ is specified as $\mathbf{x}_0\in [0,1]^n$, its final value $\mathbf{x}(T)$ is left free. Thus, the optimization each agent carries out is one of {\it free terminal condition}, \cite{kirk}.   The game (\ref{game}) lies within the framework of Theorem 6.11 in \cite{basar} and is in fact a quadratic game as we show in the Appendix so that a unique Nash equilibrium exists by Theorem 6.12 of \cite{basar}. Instead of using this result (after transforming the problem to the set up of \cite{basar}), it is easier to use the necessary conditions provided by Theorem 6.11 of \cite{basar}. We thus state those necessary conditions in the set up of our game (\ref{game}) first.   

Let $S_o$ be a trajectory or opinion space $\lbrace \mathbf{x}(t), 0 \leq t \leq T \rbrace$ and $\Gamma^i$ be a strategy space of agent $i$ so that its every mapping $\gamma^i : [0,T]\times S_o \rightarrow \Gamma^i$ is a permissible strategy for agent $i$. And define $g^i(\mathbf{x},x_0^i,u^i)$ to be the integrand of the cost functional (\ref{cost}),

{\it \textbf{Lemma 1.} For an $n$-agent dynamic game of prescribed fixed duration $[0,T]$, let

(i) \ $u^i(t)$ be continuously differentiable on $\mathbb{R}$, \ $\forall t\in [0,T]$,

(ii) $g^i(\mathbf{x},x_0^i,u^i)$ be continuously differentiable on $\mathbb{R}$, \ $\forall t\in [0,T]$, \ $i\in N$.

If $\lbrace \gamma^{i*}(t,x_0^i) = u^{i*}(t); \ i\in N \rbrace$ provides a unique open-loop Nash equilibrium solution, and $\lbrace \mathbf{x}^{*}(t), 0\leq t \leq T \rbrace$ is the corresponding opinion trajectory, then there exist $n$ costate functions $p^i(t) : [0,T] \rightarrow \mathbb{R}$, $i\in N$, such that the following relations are satisfied:
\begin{equation} \label{conditions}
	\begin{cases}
		&\dot{x}^{i*}(t) = u^{i*}(t), \\
		&\dot{p}^i(t) = -\frac{\partial {\cal{H}}^i}{\partial x^i}, \\
		&\gamma^{i*}(t,x_0^i) \equiv u^{i*}(t) = \arg \min_{u^i\in \Gamma^i} {\cal{H}}^i(p^i,\mathbf{x},x_0^i,u^i), \\
		&x^{i*}(0) = x_0^i \in [0,1], \ \ p^i(T) = 0, \ \ i\in N, 
	\end{cases}
\end{equation}
where
\begin{equation} \label{Hamiltonian}
	{\cal{H}}^i (p^i,\mathbf{x},x_0^i,u^i) = g^i (\mathbf{x},x_0^i,u^i) + p^i(t) u^i(t), \ t\in [0,T]. 
\end{equation} }
Here we note that the terminal condition of the costate functions is a consequence of the fact that the game has free terminal conditions.
Defining a Hamiltonian as in (\ref{Hamiltonian}) and using the relations in (\ref{conditions}), we can combine the state and costate equations into the following equation,
\begin{equation} \label{eqn4}
	\left[ \begin{array}{c}
	\dot{\mathbf{x}}(t) \\ \dot{\mathbf{p}}(t)
	\end{array} \right] = A
	\left[ \begin{array}{c}
	\mathbf{x}(t) \\ \mathbf{p}(t)
	\end{array} \right] + \hat{K}
	\left[ \begin{array}{c}
	\mathbf{x}(0) \\ \mathbf{p}(0)
	\end{array} \right],
\end{equation}
where
\[
	A = \left[ \begin{array}{cc} 0&-I \\ -W&0 \end{array} \right], \ \
	\hat{K} = \left[ \begin{array}{cc} 0&0 \\ K&0 \end{array} \right],
\]
where $I$ is the identity matrix of size $n$ and
$\mathbf{p}(t)=[p^1(t)\,...\,p^n(t)]', \ K=\mbox{diag} \left[k_1, ... , k_n\right]$. Here,
\[ W = \left [ \begin{array}{cccc}
q_1 & -w_{12} & \dots & -w_{1n} \\
-w_{21} & q_2   & \dots & -w_{2n}\\ 
\vdots & \vdots & \ddots & \vdots \\
-w_{n1} & -w_{n2} & \dots & q_n
\end{array} \right] , \]
where $q_i = \sum_{j\in \eta_i} w_{ij} + k_i$.
Notice that the matrix $W$ is a Laplacian-like matrix of a weighted directed graph $G$. Every $ij$-th element in the off-diagonal, $i\neq j$, shows the weight of the edge that is directed from $i$ to $j$, and the diagonal elements consist of the sum of all the weights associated with every node and its stubbornness parameter.
Solving the differential equation (\ref{eqn4}) gives,
\begin{equation} \label{eqn5}
\left[ \begin{array}{c}
\mathbf{x}(t) \\ \mathbf{p}(t)
\end{array} \right] = \left( e^{At} + \int_0^t e^{A(t-\tau)} d\tau . \hat{K} \right) \left[ \begin{array}{c}
\mathbf{x}(0) \\ \mathbf{p}(0)
\end{array} \right],
\end{equation}
where $\Phi(t) = e^{At}={\cal{L}}^{-1} \left\lbrace(sI-A)^{-1} \right\rbrace$ and $\Psi(t) = \int_0^t e^{A(t-\tau)} d\tau$. Since
\begin{equation} \label{lap}
(sI-A)^{-1} = \left[ \begin{array}{ll}
s(s^2 I - W)^{-1} & -(s^2 I - W)^{-1} \\
-W(s^2 I - W)^{-1} & s (s^2 I - W)^{-1}
\end{array} \right],
\end{equation}
one can correspondingly get the natural partitions
\[ \Phi(t) = \left[ \begin{array}{cc} \phi_{11}(t) & \phi_{12}(t) \\
\phi_{21}(t) & \phi_{22}(t) \end{array}  \right],  \ \ \
\Psi(t) = \left[ \begin{array}{cc} \psi_{11}(t) & \psi_{12}(t) \\
\psi_{21}(t) & \psi_{22}(t) \end{array}  \right].
\]
The diagonalizability assumption, although not necessary, is a simplifying assumption .

{\it \textbf{Proposition 1.} Suppose $W$ is diagonalizable so that $W = V \Lambda V^{-1}$, where $\Lambda = \mbox{diag} \left[\lambda_1, \lambda_2, ... , \lambda_n \right]$ and $V$ is the matrix whose columns are the corresponding linearly independent eigenvectors. Then, a Nash equilibrium of the game (\ref{game}) exists and is unique. The opinion trajectory of the Nash solution is given by
\begin{equation} \label{solution}
\mathbf{x}(t) = \left[ \zeta_{11}(t) - \zeta_{12}(t) \zeta_{22}^{-1}(T) \zeta_{21}(T) \right] \mathbf{x}(0),
\end{equation}
where
\begin{eqnarray*}
\zeta_{11}(t)=\phi_{11}(t) + \psi_{12}(t)K, & \ \zeta_{12}(t) = \phi_{12}(t), \\
\zeta_{21}(t)=\phi_{21}(t) + \psi_{22}(t)K, & \ \zeta_{22}(t) = \phi_{22}(t),
\end{eqnarray*}
and
\begin{eqnarray*}
\phi_{11}(t) &=& V\mbox{ diag}\left[ \pi_1 , \pi_2 , \dots , \pi_n \right]V^{-1}, \\
\phi_{12}(t) &=& -V\mbox{ diag}\left[ \hat{\pi}_1, \hat{\pi}_2, \dots , \hat{\pi}_n \right]V^{-1}, \\
\phi_{21}(t) &=& W\phi_{12}(t), \\
\phi_{22}(t) &=& \phi_{11}(t), \\
\psi_{12}(t) &=& -V\mbox{ diag}\left[ \tilde{\pi}_1, \tilde{\pi}_2, \dots , \tilde{\pi}_n \right]V^{-1}, \\
\psi_{22}(t) &=& -\phi_{12}(t),
\end{eqnarray*}
with
\begin{eqnarray*}
\pi_i = cosh \left( \sqrt{\lambda_i} \ t \right), \ \
\hat{\pi}_i = \frac{sinh \left( \sqrt{\lambda_i} \ t \right)}{\sqrt{\lambda_i}},
\\
\tilde{\pi}_i = \frac{cosh \left( \sqrt{\lambda_i} \ t \right) - 1}{\lambda_i},  \  i\in N.
\end{eqnarray*}
}

\section{Games with an Explicit Nash Solution}

The equation (\ref{solution}) in Proposition 1 will yield explicit expressions for opinion trajectories only if one can compute the eigenvalues and the eigenvectors of $W$ explicitly. In this section, we present two typical situations in which analytic expressions of the opinion trajectories are  derived.

We will say that a {\it full consensus} is reached in the network at the terminal time whenever the Nash solution of the game (\ref{game}) is such that $x^1(T)=...=x^n(T)$.  Of course, the equality may hold only for a subset of $N$, which will then indicate a partial consensus.

\subsection{Consensus in a complete information structure}

In a network where all agents are connected to each other, i.e., $\eta_i=N\setminus \{i\}$, the opinion of agent $i$ will be influenced  by all other agents and one may expect that a consensus will eventually be reached. But, due to the presence of some stubborn agents, a full consensus  may still not be reached. The present special game investigates this issue.

For simplicity and in order to get explicit solutions, we assume equal parameters for all agents, i.e., $k_i = k$, $w_{ij} = w_{ji} = w$, $\forall i \in N$ and $(i,j) \in E$.

{\bf Theorem 1.} {\it For a network of complete information structure, and where all the agents have equal parameters, the unique  Nash equilibrium is such that the opinion dynamics of agent $i$ is given by
\begin{equation} \label{solni1}
	x^i(t) = \frac{1}{n}\sum_{j=1}^n x_0^j + \gamma(t) ( x_0^i - \frac{1}{n}\sum_{j=1}^{n} x_0^j),
\end{equation}
where $\gamma(t) = \frac{k}{\lambda_1} + \left( \frac{nw}{\lambda_1} \right) \frac{cosh(\sqrt{\lambda_1}(T-t))}{cosh(\sqrt{\lambda_1}T)}$ and $\lambda_1 = k+nw$. The opinion dynamics $\mathbf{x}(t)$ with the $i$-th entry (\ref{solni1}) has the following properties:

(i) A full consensus is never achieved but the opinion dynamics will progressively converge to
\begin{equation}
	\lim_{T\rightarrow \infty}\lim_{t\rightarrow T}  x^i(t) = \frac{1}{n} \sum_{j=1}^n x_0^j + \frac{k}{\lambda_1} (x_0^i - \frac{1}{n}\sum_{j=1}^{n} x_0^j).
\end{equation}

(ii) The Nash equilibrium will be a full consensus and the opinions will converge to the average $\frac{1}{n} \sum_{j=1}^n x_0^j$ of the initial opinions if and only if there are no stubborn agents, i.e., $k=0$.

(iii) The opinion distance between any two agents at time $t\in [0,T]$ is given by}
\begin{equation}
	|\Delta x^{ij}(t)| = \gamma(t) |\Delta x_0^{ij}|,
\end{equation}
where $\Delta x^{ij}(t) = x^i(t) - x^j(t)$ and $\Delta x_0^{ij} = x_0^i - x_0^j$.

{\it Remark 1.} The opinion trajectory of every agent has two parts. The first term on the right hand side of (\ref{solni1}) is the average of initial opinions of all agents in the network, and the second term depends on the difference between the initial opinion of agent $i$ and that average. The weight of the latter is a coefficient that gets progressively closer to $k/\lambda_1$ as time passes.

{\it Remark 2.} Since we are able to derive explicit expressions for the opinion trajectories,  it is a simple matter to compute the time it takes the network to reach a consensus within $\epsilon$-vicinity of the average opinion. Or, to determine the individual influence of each parameter on the $\epsilon$-closeness to a full consensus.

{\it Remark 3.} A fast convergence to average opinion obviously requires a large $\lambda_1$, since the opinion distance as the terminal time $T\rightarrow \infty$ is
\[
	\lim_{T\rightarrow \infty} \frac{| x^i(t) - x^j(t)|}{|x_0^i - x_o^j|} = 
	\frac{k}{\lambda_1} + \frac{nw}{\lambda_1} e^{-\sqrt{\lambda_1} \ t} .
\]
The degree of closeness to full consensus at the steady state is decreased if $k\rightarrow 0$ or if $w \gg k$. A higher convergence rate requires a large $\lambda_1$, which will be the case if any one of $k$, $n$, $w$ is large. Note that in case of a larger network population, a quick consensus gets more likely because  each agent experiences more social pressure in a complete information graph topology.

\subsection{Consensus under a leader}

The leader (agent $1$) in this network can be considered as some political analyst who can influence the opinions of other agents through electronic media. Therefore, the leader can influence the opinions of other agents, but not the other way round, based on the value of their influence and stubbornness parameters. Due to that influence, they tend to adjust their opinions closer to leader's opinion. The network is represented by a directed graph where the edges are directed from agents towards the leader. Thus $\eta_1=\emptyset, \eta_i=\{1\}, \forall i\in N\setminus \{1\}$. It follows that 
in this special game (\ref{game}), $w_{ij}\neq 0$ only if $j=1$.

The question we investigate is whether the leader's opinion will prevail under all parameter values given enough time.
One of course expects that a full consensus may not be achieved in a finite duration whenever stubborn agents exist but if some agent $i$ is not stubborn, i.e., $k_i=0$, then that agent will make consensus with the leader.

{\bf Theorem 2.} {\it For a network in which all agents are unilaterally connected to the leader (agent $1$), the unique Nash equilibrium is such that the opinion dynamics of agents are given by 
\begin{equation} \label{solni2}
	\begin{cases}	
		\!\begin{aligned}		
		x^1(t) = & x_0^1, \\
		x^i(t) = & \frac{k_i x_0^i + w_{i1} x_0^1}{\lambda_i} + \xi_i(t) \left( x_0^i - x_0^1 \right),
		\end{aligned}
	\end{cases}
\end{equation}
where $\xi_i(t) = \left( \frac{w_{i1}}{\lambda_i} \right)  \frac{cosh\left(\sqrt{\lambda_i}(T-t)\right)}{cosh(\sqrt{\lambda_i}T)}$ and $\lambda_i = k_i+w_{i1}$, $\forall i \in N\setminus \lbrace 1\rbrace$. The opinion dynamics $\mathbf{x}(t)$ with the $i$-th entry (\ref{solni1}) has the following properties:

(i) The leader never changes its initial opinion and the opinions of other agents $i \in N \setminus \lbrace 1 \rbrace$, converge to
\begin{equation}
\lim_{T\rightarrow \infty}\lim_{t\rightarrow T}  x_i(t) =  \frac{k_i x_0^i + w_{i1} x_0^1}{\lambda_i}.
\end{equation}

(ii) For  $i \in N \setminus \lbrace 1 \rbrace$, opinion of agent $i$ will converge to the leader's opinion as $T\rightarrow\infty$ if and only if $k_i = 0$. 

(iii)\ The opinion distance of any agent to the leader is given by
\begin{equation}
	|\Delta x^{i1}(t)| = \left( \frac{k_i}{\lambda_i} + \xi_i(t) \right) |\Delta	x_0^{i1}|.
\end{equation} }
where $\Delta x^{i1}(t) = x^i(t)-x^1(t)$ and $\Delta x_0^{i1}=x_0^i - x_0^1$.

{\it Remark 4.} Note that the consensus in the long run is a convex combination of the initial opinions of agent $i$ and the leader. In this convex rivalry, a stubborn agent will stand alone.

{\it Remark 5.} It is possible to determine the time in which agent $i$ is $\epsilon$-close to the opinion maintained by the leader. Similarly, it is straightforward to examine the sensitivity of an $\epsilon$-consensus to each parameter value $w_{i1}, k_{i}$.

{\it Remark 6.} A fast convergence to the opinion of the leader requires a large $\lambda_i$. This will be the case if the value of $w_{i1}$ or $k_i$ is large. Although the convergence time is increased, the property (i) shows that final value of the opinion will incline towards either the leader's or agent's initial opinion depending on whether $w_{i1}>k_i$ or $w_{i1}<k_i$, respectively.

\section{Some Simulations of the General Opinion Dynamics Game}

\begin{figure}[h]
	\centering
	\begin{subfigure}[b]{0.5\textwidth}
		\centering
		\includegraphics[width=\textwidth, trim={0 15mm 0 10mm},clip]{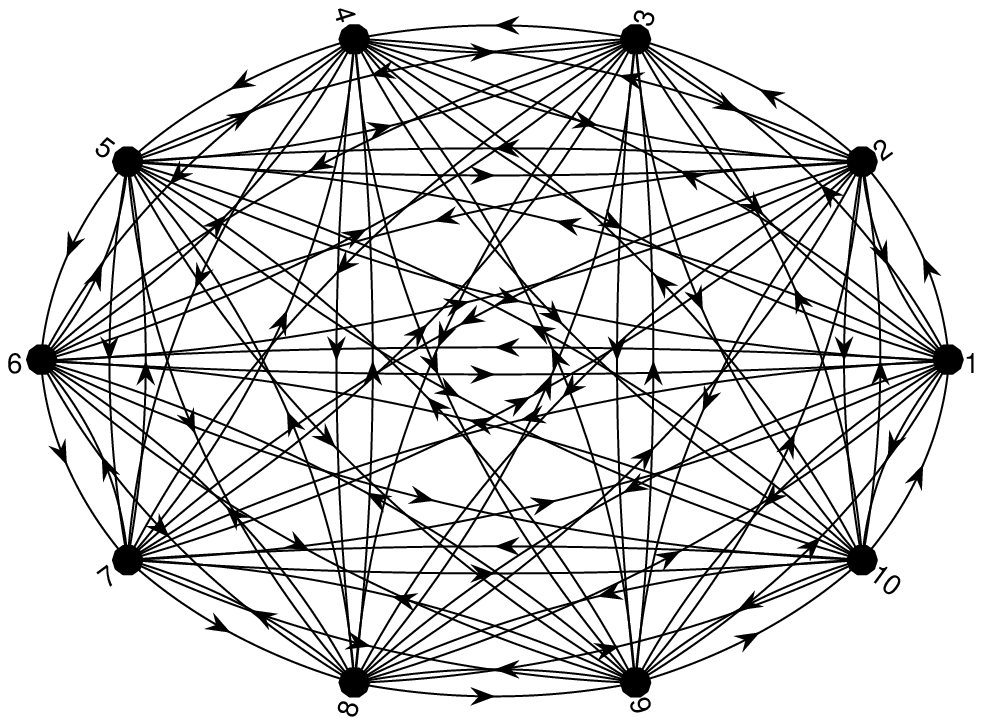}
		\caption{}
		\label{Fig1:G1}
	\end{subfigure}
	\begin{subfigure}[b]{0.35\textwidth}
		\centering
		\includegraphics[width=\textwidth]{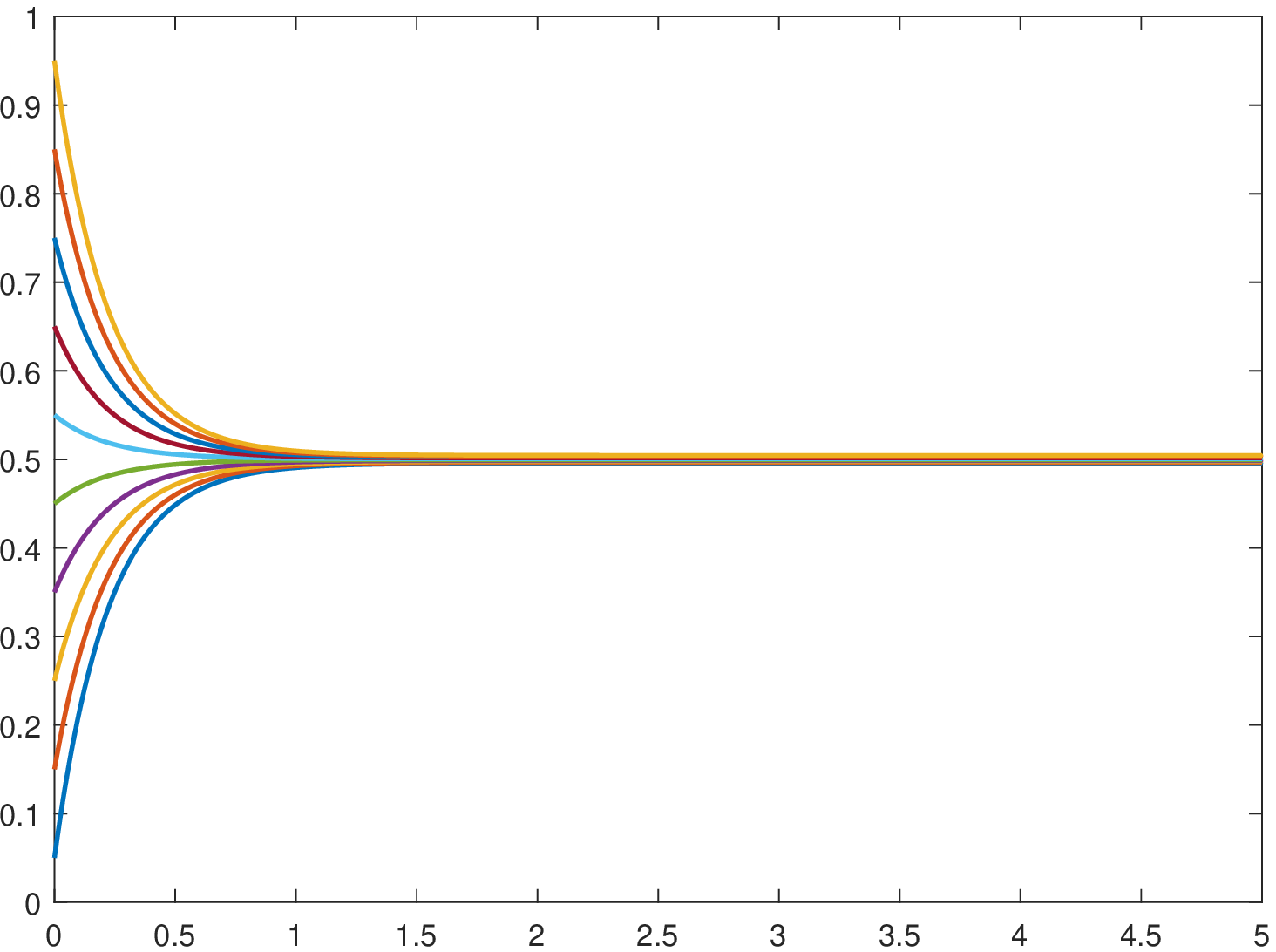}
		\caption{}
		\label{Fig1:01}
	\end{subfigure}
	\begin{subfigure}[b]{0.35\textwidth}
		\centering
		\includegraphics[width=\textwidth]{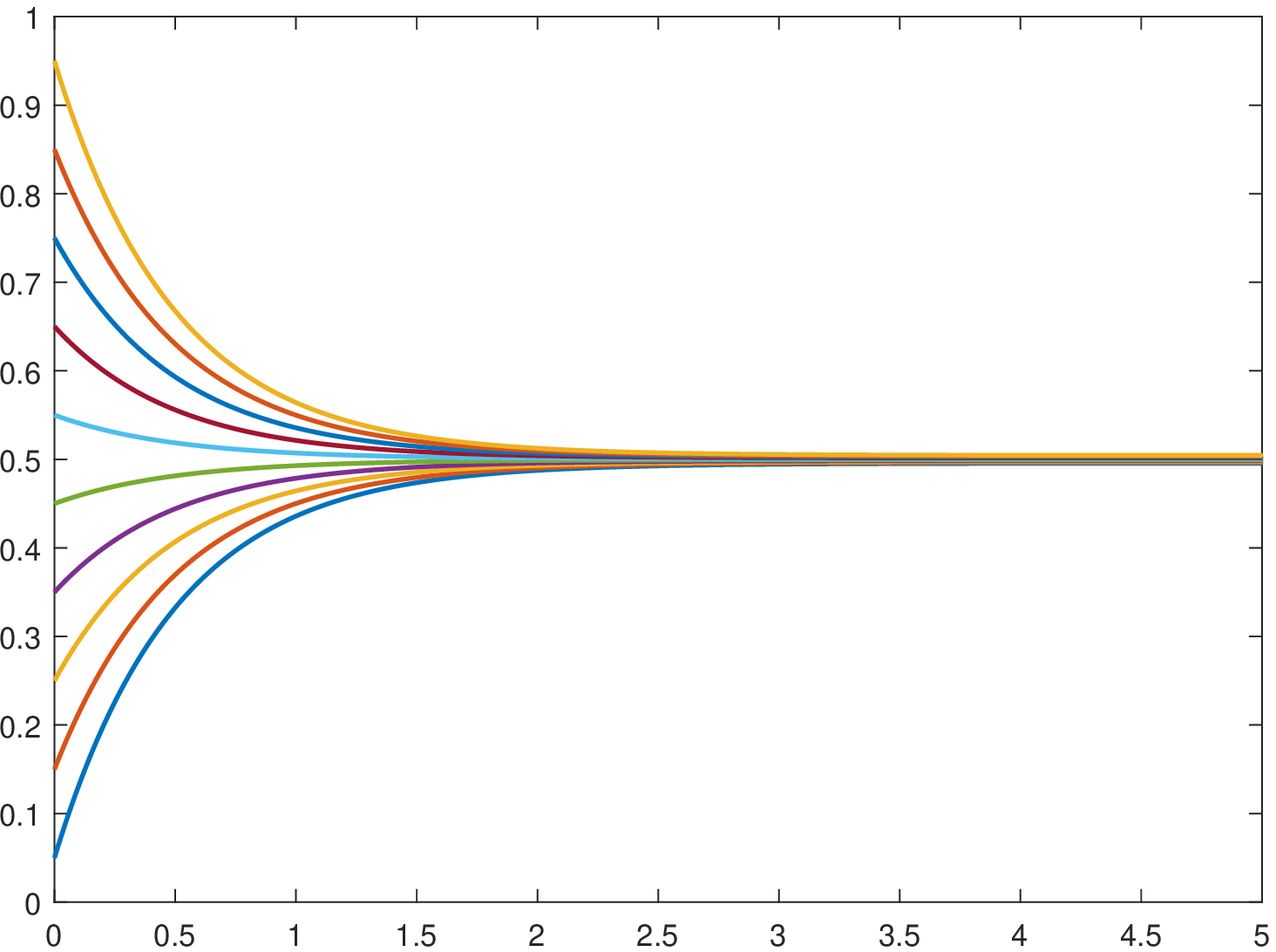}
		\caption{}
		\label{Fig1:02}
	\end{subfigure}

	\caption{Complete Information Structure}
	\label{Fig1}
\end{figure}
\begin{figure}[h]
	\centering
	\begin{subfigure}[b]{0.5\textwidth}
		\centering
		\includegraphics[width=\textwidth, trim={0 15mm 0 10mm},clip]{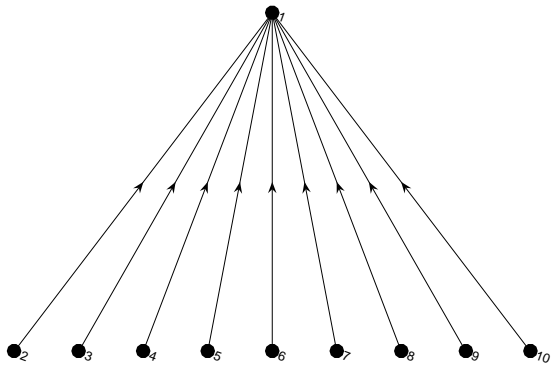}
		\caption{}
		\label{Fig2:G2}
	\end{subfigure}
	\begin{subfigure}[b]{0.35\textwidth}
		\centering
		\includegraphics[width=\textwidth]{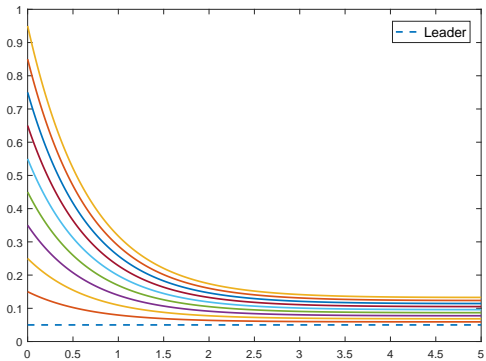}
		\caption{}
		\label{Fig2:01}
	\end{subfigure}
	\begin{subfigure}[b]{0.35\textwidth}
		\centering
		\includegraphics[width=\textwidth]{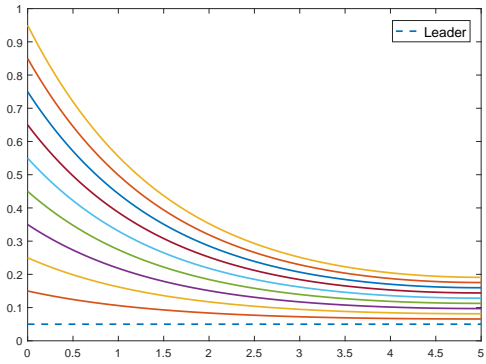}
		\caption{}
		\label{Fig2:02}
	\end{subfigure}
	\caption{A network with one leader (agent 1)}
	\label{Fig2}
\end{figure}
\begin{figure}[h]
	\centering
	\begin{subfigure}[b]{0.5\textwidth}
		\centering
		\includegraphics[width=\textwidth, trim={0 15mm 0 10mm},clip]{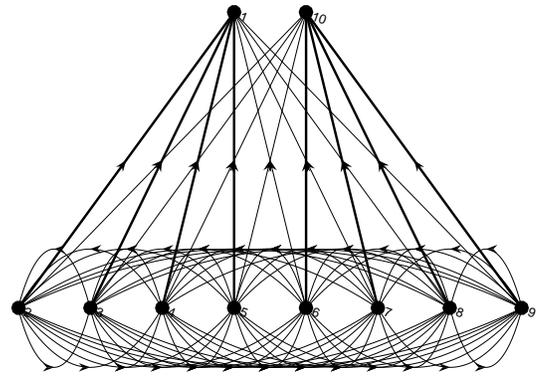}
		\caption{}
		\label{Fig3:G3}
	\end{subfigure}
	\begin{subfigure}[b]{0.35\textwidth}
		\centering
		\includegraphics[width=\textwidth]{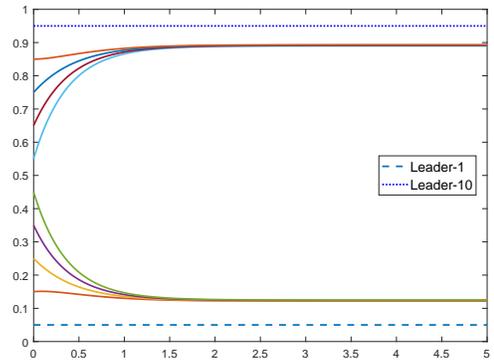}
		\caption{}
		\label{Fig3:01}
	\end{subfigure}
	\begin{subfigure}[b]{0.35\textwidth}
		\centering
		\includegraphics[width=\textwidth]{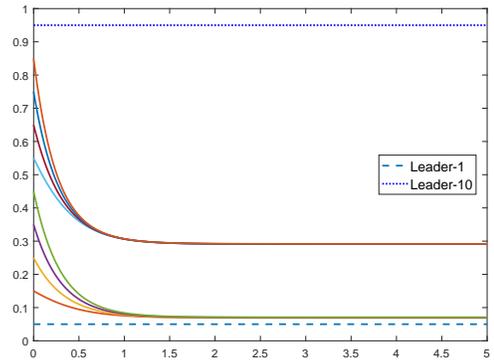}
		\caption{}
		\label{Fig3:02}
	\end{subfigure}
	\caption{A network with two leaders (agent 1 and 10)}
	\label{Fig3}
\end{figure}

We have simulated a number of network structures, focusing on those with diagonalizable W matrix, and investigated the effect of some parameters on opinion dynamics. Due to space limitations, we present three simulations that illustrate in Figures 1, 2, and 3 the results of Theorems 1, 2 and Proposition 1. Here, we confine our investigation of parameter effects to only see what happens if the control term in the cost function is dominant or not dominant. The simulation results in Figures 1 and 2 coincide with the plots obtained by the analytic expressions of the opinion trajectories from Theorems 1 and 2. In all simulations, number of agents is $n=10$ and terminal time is $T=5$ units. Initial opinion levels of the agents are chosen as $\mathbf{x}(0) = [0.05, 0.15, 0.25, 0.35, 0.45, 0.55, 0.65, 0.75, 0.85, 0.95]'$.

Fig. 1(a) illustrates  a complete information structure of Theorem 1 and Fig. 2(a), the one leader network of Theorem 2, respectively. In Fig. 1(b) and Fig. 2(b), $w=2$ and $k=0.2$ for all agents. In Fig. 1(c) and Fig. 2(c), we set $w=0.4$ and $k=0.04$. The reduction of the weights of both the influence and the stubbornness parameters have the effect of bringing forth the penalization of the control term in the cost function. This results in slowing down the convergence rate although the opinions converge to the same values in both cases.

To illustrate the result of Proposition 1, we present a 2-leader network (Fig. 3(a)) in which agent 1 and agent 10 are two leaders. The opinion trajectories are obtained via the expression (\ref{solution}) after computing $\Phi(t)$ and $\Psi(t)$ through MATLAB. It is assumed that half of the followers support each leader. The followers of leader-1 can be named as followers-1, and of leader-10 as followers-10. The followers also have influence among themselves in a society, of course, but an agent can be assumed to have more impact on his fellow supporters. We set $k_i=0.2 \ \forall i\in N$. The matrix $W$ in (\ref{eqn4}) is such that $w_{1i}=w_{ni}=0, \ \forall i\in N$ and $n=10$. In Fig. 3(b), we assume that the influences of leader-1 and leader-10 on their followers is 10; the social impact of followers-1 and followers-10 among themselves is 2; the cross impact of followers-1 on followers-10, and vice versa, is 0.2; and the influence that followers take from other leader is 100 times less than their own leader's influence. In this case, agents follow their respective leaders. However, if we assume that followers-1 are more loyal to their leader and they run a good campaign in order to attract the followers-10, then they are able to steal followers-10 from their leader. For Fig. 3(c), we increase the influence of leader-1 to 20 and the cross impact of followers-1 on followers-10 to 10, while all other parameters are the same. It can be seen that rather than following leader-10, followers-10 tend to follow followers-1 due to social impact.

\section{Conclusion}

The main conclusion of this study is that a consensus, Nash equilibrium, can spontaneously be reached from independent motives of agents in a social network in the long run. How unanimous this consensus is depends on the initial differences of opinion, on the susceptibility of agents to influence, and on the stubbornness of agents. If one member is singled out as a leader who firmly sticks to its opinion, then a consensus about the leader's opinion is again formed in the long run. If an opinion game is played in a finite interval a full consensus is never reached in the presence of stubborn members.

The game with leader can also be viewed as a game of learning in which the leader is the teacher and the others, students. Here, it may be more instructive to examine the situation when learning is poorly achieved.  It is clear that initial ignorance of the subject and reluctance to learn, all contribute to this. But, a low willingness to update is also a negative factor.

The game studied can be extended in several directions. The opinion on a single issue is not essential and one can consider each agent having opinions on several issues as e.g., in \cite{krause}. The technicality such an extension requires is that $x^i(t)$ is no longer a scalar but a vector with each entry representing the level of opinion on one issue. One can also extend the game considered here to a network in which agents can have different ``types" of motives. Then, the integrands of the cost functions the agents use will not have a uniform structure, e.g., non-quadratic cost functions may be employed along with quadratic ones.

\section*{Appendix}

\subsection*{A1. Proof of Proposition 1}

We first note that the cost functional $L^i$ of (\ref{cost}) can be transformed to a quadratic functional by  
\[ 
L^i(\mathbf{z}^i, u^i) = \frac{1}{2} \int_0^T \left({\mathbf{z}^i}(t)' G^i \mathbf{z}^i(t)+[u^{i}(t)]^2\right)\ dt,
\]
where
\[ 
\mathbf{z}^i = \left[ \Delta x^{i1}, ... , \Delta x^{i \ i-1}, \Delta x^{i \ i+1}, ... , \Delta x^{in}, x^i-x_0^i \right]',
\]
%\mathbf{z}^i = \left[ x^i-x^1, ... , x^i-x^{i-1}, x^i-x^{i+1}, ... , x^i-x^n, x^i-x_0^i \right]',
and
\[ G^i = \mbox{diag} \left[ w_{i1}, ... , w_{i \ i-1}, w_{i \ i+1}, ... , w_{in}, k_i \right] \geq 0.
\]
This fact allows one to employ Theorem 6.12 of \cite{basar} and the opinion trajectories in a unique Nash solution can be derived. However, because the transformation above is not a simple one, it is much easier to directly obtain the Nash solutions through the necessary conditions of Lemma 1. This is the approach used in this Appendix. The uniqueness of a Nash solution when one exists is, however, a direct consequence of the above transformation and will not be separately addressed.

Let us write (\ref{eqn5}) as
\begin{equation} \label{eqn9}
\left[ \begin{array}{c}
\mathbf{x}(t) \\ \mathbf{p}(t)
\end{array} \right] = \left[ \begin{array}{cc}
\zeta_{11}(t) & \zeta_{12}(t) \\ \zeta_{21}(t) & \zeta_{22}(t)
\end{array} \right] \left[ \begin{array}{c}
\mathbf{x}(0) \\ \mathbf{p}(0)
\end{array} \right],
\end{equation}
and note that the expressions for $\Phi(t)$, $\Psi(t)$, and their partitions are obtained by the inverse Laplace transform of (\ref{lap}) via a matrix partial fraction expansion (see e.g. Lemma A.3  in \cite{ozguler2} for a similar procedure).  By Lemma 1, $\mathbf{p}(T) = 0$ so that from equation (\ref{eqn9}) evaluated at $t=T$ we get $\mathbf{p}(T)=\zeta_{21}(T)\mathbf{x}(0) + \zeta_{22}(T)\mathbf{p}(0) = 0$. Since, by its expression in Proposition 1, $\phi_{22}(T)=\zeta_{22}(T)$ is nonsingular, we obtain 
$\mathbf{p}(0)=- \zeta_{22}^{-1}(T) \zeta_{21}(T) \mathbf{x}(0)$. Substituting into 
(\ref{eqn9}), the solution (\ref{solution}) is obtained. 

\subsection*{A2. Proof of Theorem 1}
We have  $K = kI$ and $W = qI - w({\cal{I}}-I)$; where $q = w(n-1) + k$, $I$ is the identity matrix and $\cal{I}$ is the matrix of all ones. For a matrix $W$, we can easily find the eigenvalues and the eigenvectors, \cite{ozguler}. They are 
$\lambda_1 = q+w= k+nw$ with multiplicity $n-1$ and $\lambda_2 = q+w-nw= k$ with multiplicity 1.

Computing the corresponding eigenvectors, 
we obtain $W = V \Lambda V^{-1}$, where $\Lambda = \mbox{diag} \left[ \lambda_1, ... ,\lambda_1, \lambda_2 \right]$, $V = \left[ \begin{array}{cc} \hat{V}_1&\hat{V}_2 \end{array} \right]$, and $V^{-1} = \left[ \begin{array}{c} \tilde{V}_1 \\
\tilde{V}_2 \end{array} \right]$, 
where $\hat{V}_{2}$ and $\tilde{V}_2$ are given by 
\[
 \hat{V}_2 = \left[ \begin{array}{c}1 \\ \vdots \\ 1\end{array} \right]_{n\times 1},\;\tilde{V}_2 = \frac{1}{n}\left[ \begin{array}{ccc}
1&\dots&1
\end{array} \right]_{1\times n},
\]
because they are, respectively, right and left eigenvectors of $O$ associated with
$\lambda_{2}$. 
Since $\hat{V}_1 \tilde{V}_1 + \hat{V}_2 \tilde{V}_2 = I$, we have  $\hat{V}_2 \tilde{V}_2 = \frac{1}{n} {\cal I}$ and $\hat{V}_1 \tilde{V}_1=I-\frac{1}{n} {\cal I}$. Also note that, in the notation of Proposition 1,  $\lambda_i=\lambda_{1}$ for $i=1,...,n-1$ and 
$\lambda_n=\lambda_{2}$. This significantly simplifies the expressions for $\phi_{ij}(t)$ and $\psi_{ij}(t)$ of Proposition 1. For instance,
\[
\phi_{11}(t)= \left[ \begin{array}{cc} \hat{V}_1&\hat{V}_2 \end{array} \right]\left[ \begin{array}{cc}\alpha I&0\\0&\beta\end{array} \right]\left[ \begin{array}{c} \tilde{V}_1 \\
\tilde{V}_2 \end{array} \right]=\alpha I+\frac{1}{n}(\beta-\alpha){\cal I},
\]
where $\alpha = cosh(\sqrt{\lambda_{1}}t), \beta = cosh(\sqrt{\lambda_{2}}t)$. Thus, $\phi_{11}(t)$ is a matrix with diagonal entries all equal to $\frac{1}{n}(\beta+(n-1)\alpha)$ and off-diagonal entries all equal to $\frac{1}{n}(\beta-\alpha)$.  Simplifying all partition matrices of Proposition 1 with this procedure and substituting in (\ref{solution}), one arrives at 
\begin{equation} \label{soln2}
	\mathbf{x}(t) = \frac{1}{n} \biggl( [ 1+(n-1)\gamma(t) ] I + ( 1-\gamma(t) ) ({\cal{I}} - I) \biggr) \mathbf{x}(0),
\end{equation}
where $\gamma(t) = \frac{k}{\lambda_1} + \left( \frac{nw}{\lambda_1} \right) \frac{cosh(\sqrt{\lambda_1}(T-t)}{cosh(\sqrt{\lambda_1}T)}$.
The $i$-th row of the right hand side of (\ref{soln2}) simplifies to the right hand side of (\ref{solni1}).

\subsection*{A3. Proof of Theorem 2}
Given the  information structure of this game, we get a lower triangular matrix
\[ W = \left[ \begin{array}{ccccc}
q_1 &  & & & \\
-w_{21} & q_2 &  && \\
-w_{31} & 0 & q_{3} && \\
\vdots& \vdots & \ddots &\ddots &\\
-w_{n1} & 0 & \dots& 0 & q_n
\end{array} \right], \]
where $q_1 = k_1$, $q_i = k_i + w_{i1}$ $\forall i \in N \setminus \{1\}$. It turns out that $W$ is diagonalizable with  $W = V \Lambda V^{-1}$. Here, $\Lambda = \mbox{diag}[q_1, ... , q_n]$ and the matrix $V$ and its inverse are lower triangular in the form 
\[ V(v_{i1}) := \left[ \begin{array}{ccccc}
1 &  & & &\\
v_{21} & 1  & & & \\
v_{31} & 0 & 1 &  \\
\vdots &\vdots & \ddots&\ddots& \\
v_{n1} & 0 & \dots & 0 & 1
\end{array} \right].
\]
where $V=V(\nu_{i1})$, $V^{-1} = V(-\nu_{i1})$ with $\nu_{i1} = \frac{w_{i1}}{q_i - q_1}$ $\forall i \in N\setminus \{1\}$. In the notation of Proposition 1, $\lambda_{i}=q_i, \ \forall i\in N$. Also exploiting the common structure
$
\left[ \begin{array}{cc}*&0\\ *&Q\,\mbox{or}\,I\end{array}\right]
$
of the matrices $W, V, V^{-1}$, where $Q = \mbox{diag}[q_2,...,q_n]$, the matrices $\phi_{ij}$ and $\psi_{ij}$ of Proposition 1 can all be simplified and (\ref{solution}) can be found as
\begin{equation} \label{eqn18}
\mathbf{x}(t) = \left[ \begin{array}{ccccc}
1 & & & & \\
\rho_2(t) & \sigma_2(t) & & & \\
\rho_3(t) & 0 & \sigma_3(t) & & \\
\vdots & \vdots & \ddots & \ddots & \\
\rho_n(t) & 0 & \dots & 0 & \sigma_n(t)
\end{array} \right] \mathbf{x}(0),
\end{equation}
where
\[ \rho_i(t) = \frac{w_{i1}}{q_i}  - \xi_i(t),
\]
and
\[ \sigma_i(t) = \frac{k_i}{q_i}  + \xi_i(t).
\]
The right hand side of (\ref{solni2}) is obtained by simplifying the $i$th row of (\ref{eqn18}).



%\bibliographystyle{IEEEtran}

\end{document}